\documentclass[epj,twocolumn]{webofc}%
\usepackage{graphics}
\usepackage[varg]{txfonts}   
\woctitle{ISVHECRI 2016}
\begin{document}
\title{Challenges for Cosmic-ray Experiments}
%
%

\author{Thomas Gaisser\inst{1}\fnsep\thanks{\email{gaisser@bartol.udel.edu
    }} 
}

\institute{Bartol Research Institute and Department of Physics and Astronomy\\
University of Delaware\\
Newark, DE 19716 USA     }

\abstract{%
This paper is a commentary on presentations at ISVHECRI 2016 related to cosmic-rays, gamma-rays
and neutrinos.  Its goal is to highlight the unanswered questions raised during the conference about
the sources of these cosmic particles and the relations among them.
}
\maketitle
\section{Introduction}
\label{intro}
A current theme of astro-particle physics is "multi-messenger astronomy," which
emphasizes use of gamma-rays and neutrinos to address still unanswered questions
on the origin of high-energy cosmic rays.  With this theme in mind, I review the
presentations on cosmic-ray spectra and composition, gamma-ray astronomy and
neutrino astronomy presented at ISVHECRI 2016.  The role of hadronic interactions is
addressed in the review paper of Tanguy Pierog~\cite{Pierog}.

\section{Some cosmic-ray questions}
\label{sec-1}

\begin{figure}[!t]
\centering
\includegraphics[width=7cm,clip]{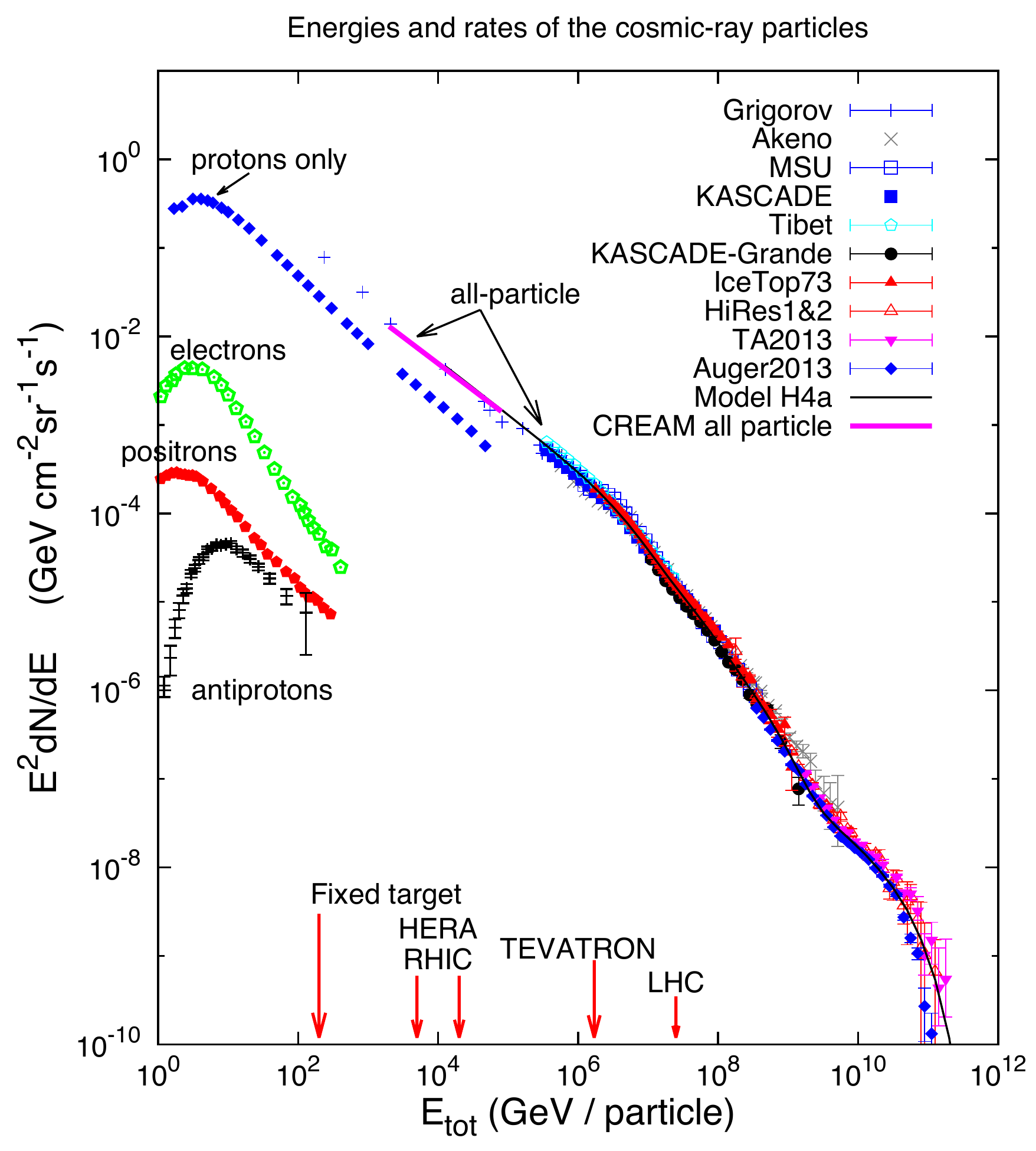}
\caption{Overview of the energy spectra of various components of the cosmic radiation
(Fig. 2.1 of~\cite{book}).}
\label{fig:GER2_1}       
\end{figure}

Figure~\ref{fig:GER2_1} gives a global view of the cosmic-ray
spectrum.
The abundant elements of the primary cosmic-ray spectrum are measured
accurately to energies higher than a TeV per nucleon
with spectrometers in space~\cite{Adriani:2011cu,Aguilar:2015ooa,Aguilar:2015ctt}.  
Calorimetric measurements with balloon-borne detectors~\cite{Panov:2011ak,Ahn:2009tb}
extend direct measurements to higher energy but with
somewhat less precision.  This means that we have good coverage of the
composition with  direct measurements up to about $100$~TeV energy per nucleus.
Indirect measurements with large detectors on the surface are needed for the
higher energy cosmic rays.  There are several questions of current interest
associated with the various features in the energy spectrum:  
\begin{itemize}
\item What is the composition
in the knee region and how does it connect with direct measurements at lower energy?
\item What is the cause of the hardening of the spectrum around 20 PeV?
\item Where is the transition from Galactic to extra-galactic cosmic rays and how is it related to 
composition around the ankle?
\item What is responsible for the apparent end of the spectrum around 100 EeV?
\item Does the difference between the Auger and the TA spectrum in the cutoff region
show that the cosmic-ray spectrum is different in different regions of the sky?
\end{itemize}

\subsection{The knee region}

Most air shower measurements have a threshold around a PeV, while direct
measurements extend only to $\sim 100$~TeV.  An exception is the ARGO-YBJ
RPC carpet detector at high altitude in Tibet, which has the potential to
cover the gap between direct and indirect measurements~\cite{diSciascio}.
TAIGA-HISCORE is also starting to measure the spectrum down to $\sim\,300$~TeV~\cite{Kuzmichev}.
ARGO-YBJ have reported measurements  of  both the all-particle spectrum
and the spectrum of the light (p + He) component~\cite{Bartoli:2015vca}.  The
measurements cover the energy range from $\sim 10$~TeV to the knee region.
While the measurement of the all-particle spectrum agrees with several other
EAS measurements through the knee region, the light component appears to steepen
starting around $700$~TeV~\cite{diSciascio,Bartoli:2015vca}.
In contrast, KASCADE~\cite{Antoni:2005wq,Haungs} shows the proton steepening above a PeV.
In his presentation, DiSciascio compares the ARGJO-YBJ result
with the H\"{o}randel parameterization~\cite{Hoerandel:2002yg} of p+He, 
which also steepens at higher energy.  The IceCube/IceTop composition analysis~\cite{Aartsen:2015awa}
starts around 3~PeV, too high to provide insight on this question.

\begin{figure}[!t]
\centering
\includegraphics[width=7cm,clip]{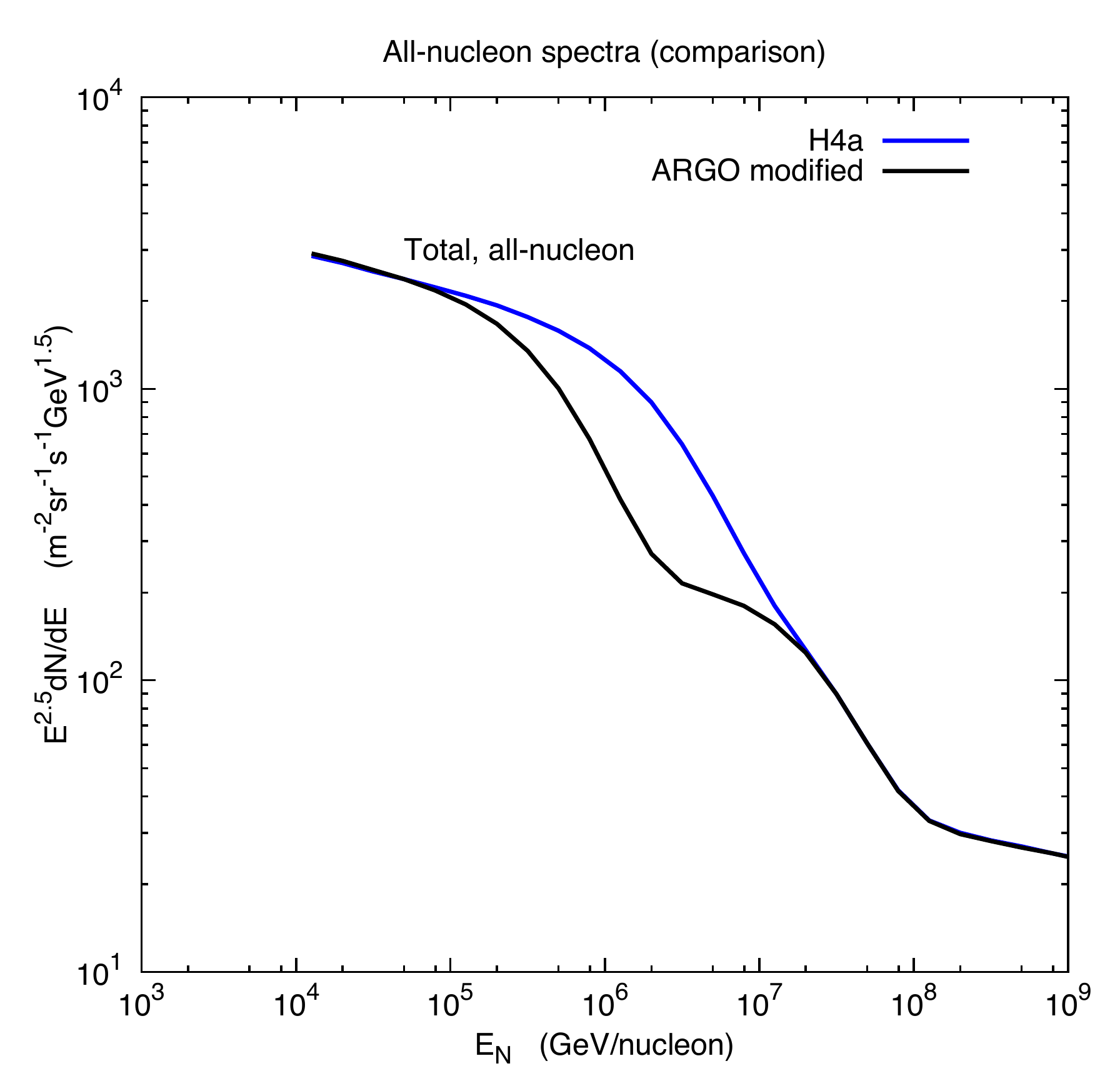}
\caption{The spectrum of nucleons for the H4a model~\cite{Gaisser:2011cc} compared with a 
modified version in which the cutoff rigidities for p and HE are reduced to 700~GV
and the all-particle spectrum is restored by increasing the contribution of
the CNO and Fe groups.}
\label{fig:compare}       
\end{figure}

\begin{figure*}
\centering
\includegraphics[width=7cm,clip]{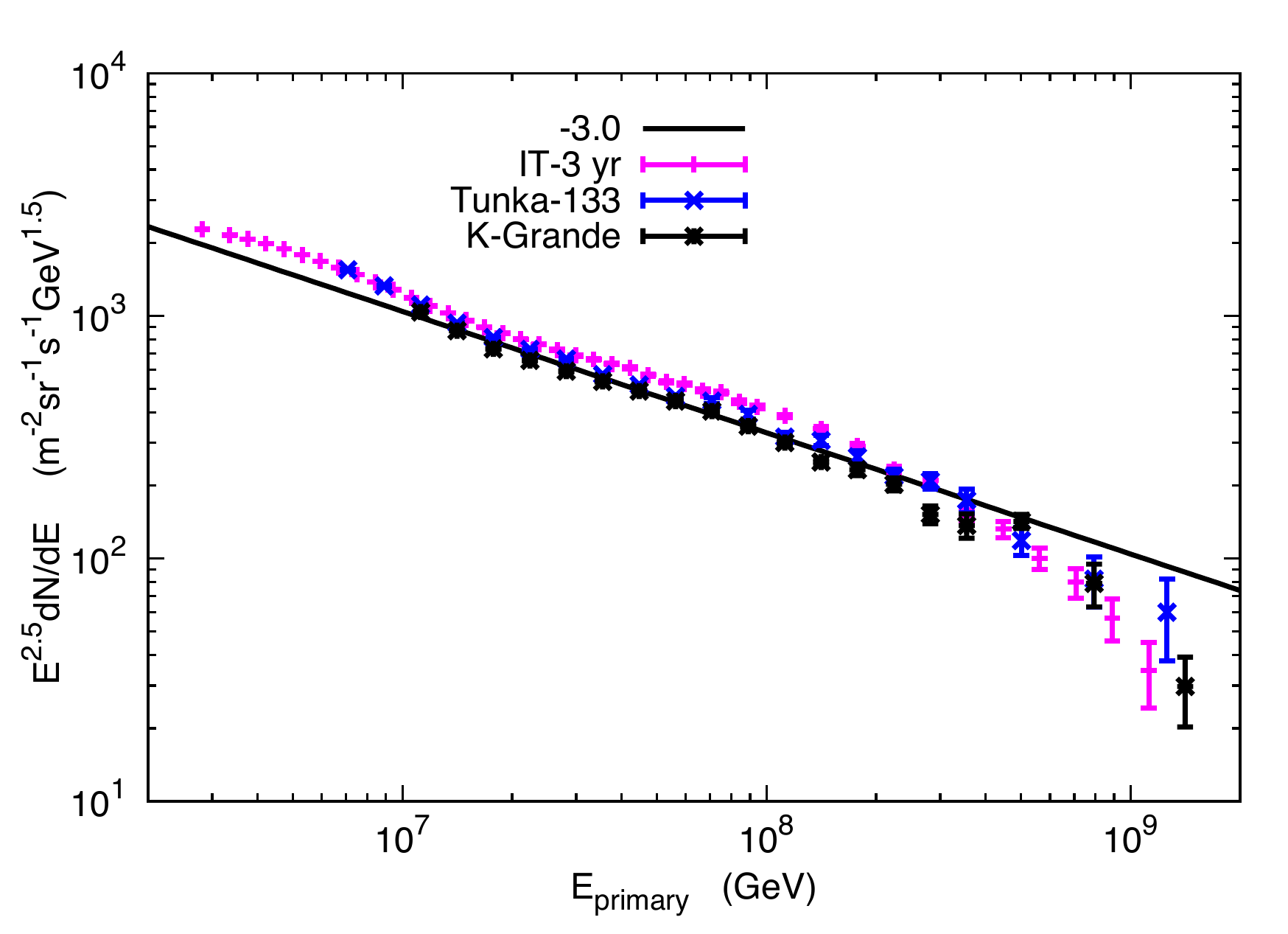},\includegraphics[width=7cm,clip]{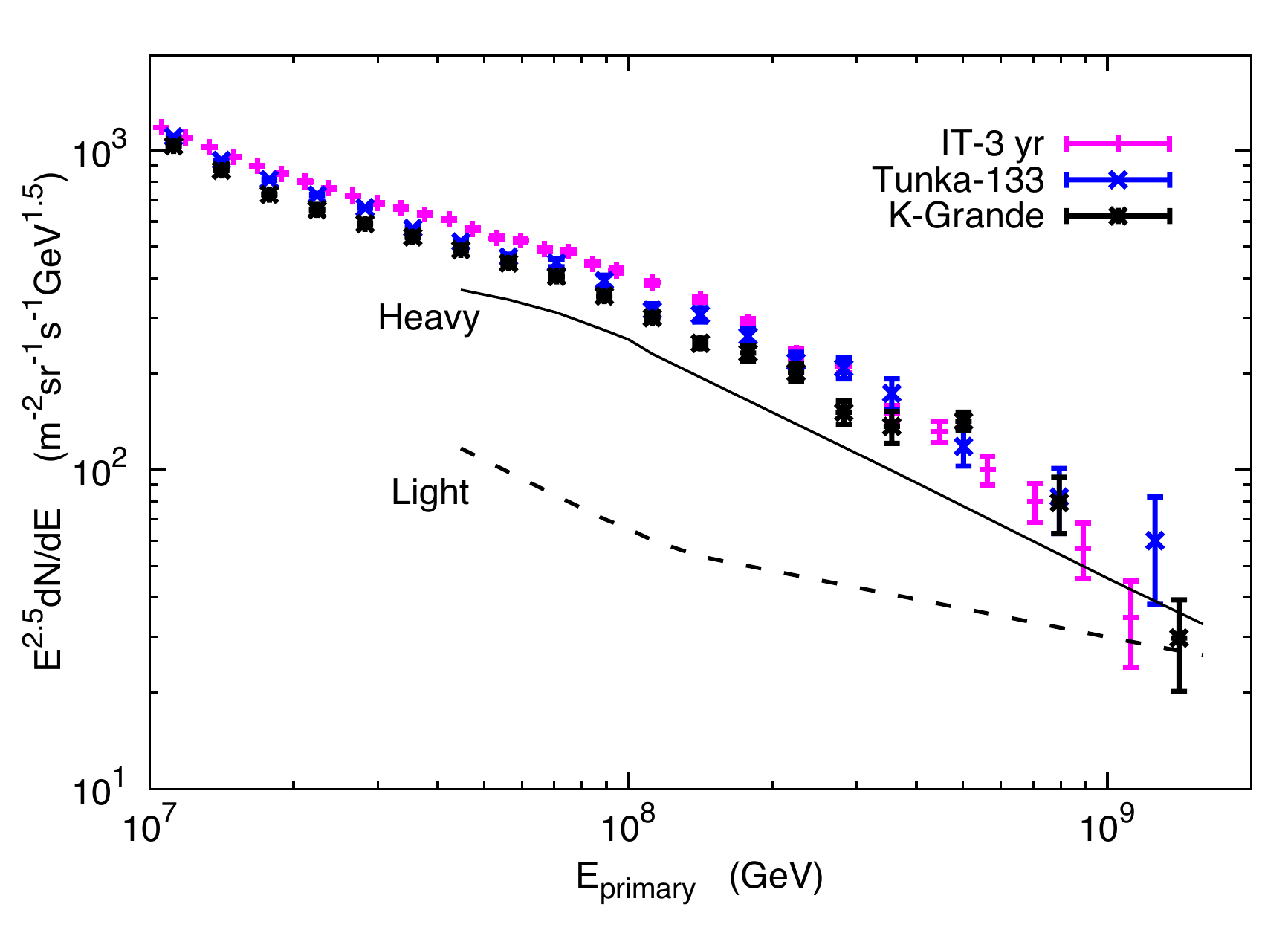}
\caption{Measurements of the spectrum between the knee and the ankle, Left: compared
to a constant $E^{-3}$ differential spectrum (Fig. 17.2 of~\cite{book}) and Right:
showing  the heavy and light fractions
as measured by KASCADE-Grande~\cite{Apel:2013ura} separately (Fig. 17.3 of~\cite{book}).}
\label{fig:GER17}
\end{figure*}

Indirect measurements of the composition with EAS detectors are difficult, and the
ARGO-YBJ result points to an ambiguity that needs to be resolved.  A practical
aspect is its implication for the atmospheric neutrino
flux at high energy relevant for IceCube.   Calculation of the flux of atmospheric
neutrinos depends on the spectrum of nucleons as a function of energy per nucleon,
which is dominated by protons and helium.  If the proton and helium components
steepen at 700~GV, then there should be a compensating increase in heavier nuclei
to keep the all-particle spectrum constant.  The sketch in Fig.~\ref{fig:compare}
illustrates the effect, which would likely be a suppression of the flux of
nucleons in a range around a PeV that arises if the all-particle spectrum is dominated
by heavy nuclei in this region.  This in turn would significantly reduce the
flux of muons and muon-neutrinos around 100~TeV.

\subsection{From the knee to the ankle}

Figure~\ref{fig:GER17} (left) compares measurements of the spectrum by 
KASCADE-Grande~\cite{Haungs,Apel:2012tda}, TUNKA-133~\cite{Prosin,Prosin:2014dxa} and
IceTop~\cite{Dembinski,Aartsen:2015awa,Aartsen:2013wda}.  The solid line shows a spectrum
with a constant differential index of $-3$.  The data show clear structure
between the knee and the ankle, with a hardening around $2\times 10^{16}$~eV
and a second knee above $10^{17}$~eV.  The KASCADE-Grande analysis 
uses the fraction of muons to separate the spectrum into light and heavy
components~\cite{Haungs,Apel:2013ura}. The data suggest that the sub-dominant light component
increases relative to the heavy component as energy increases toward the PeV region,
as shown in Fig.~\ref{fig:GER17} (right).  A possible interpretation is that the increase
of the light component reflects the population of cosmic rays from extragalactic
sources, while the steeper heavy component is the end of the Galactic population.

\subsection{The highest energy cosmic rays}

Measurements of the cosmic-ray spectrum to the highest energy were presented
at ISVHECRI 2016 from both Telescope Array (TA)~\cite{Matthews} and Auger~\cite{Martello}.
They are in excellent agreement with each other through the ankle region within their
systematic uncertainties in energy.  However, after shifting the Auger energy assignment up
by 8\% (or the TA spectrum down by a similar amount) the TA spectrum remains somewhat higher
than the Auger spectrum above 10~EeV.

The question of composition of the highest energy cosmic rays has long been an
important unresolved issue.  Both TA and Auger find a large fraction of protons
in the EeV range, above which the interpretations differed, with Auger preferring
heavier and TA lighter composition.  Inferences about composition are based  on both the mean depth
of shower maximum as a function of energy and on fluctuations in depth of 
maximum in each energy bin, and they depend on the hadronic
interaction model used to make the interpretation.  The TA presentation~\cite{Matthews} 
includes a plot of mean depth of maximum for both experiments obtained by the joint
composition working group that includes members of both experiments.  The results of the
two experiments are not inconsistent with each other.  In comparison with
the interaction model QGSJETII-03 the depth of maximum measurements are between
protons and iron, but closer to the proton limit.  Thus at present the composition at
the highest energy remains an open question.

Composition from 1-100~EeV is the key to what is one of the most important open
questions in cosmic-ray physics, namely, the cause of the apparent cutoff in the
spectrum at 100 EeV.  There are two possibilities.  If protons dominate at high energy,
the natural interpretation would be the GZK process~\cite{Greisen:1966jv,Zatsepin:1966jv}, energy loss
to photo-pion production
during propagation in the cosmic microwave background radiation.
The other possibility is that the accelerators are reaching their
maximum rigidity, as suggested by the Hillas plot~\cite{Hillas:1985is}.
The Auger presentation~\cite{Martello} illustrates the different
energy-dependent compositions that characterize each of these possibilities~\cite{diMatteo}.
The GZK explanation requires mostly protons at the highest energy
while in the Hillas case an increasing fraction of heavy nuclei
would be expected as the cutoff in rigidity affects protons first.
In both cases effects of nuclear fragmentation during propagation
must accounted for in addition to the source composition.

Both TA and Auger have initiated upgrades aimed at understanding the
composition and the related question of anisotropy of the highest energy
cosmic rays.  TA$\times 4$ will expand the surface detector with two
large scintillator arrays (grid spacing 2.08~km) adjacent to
the present array.  One of the main aims is to get a better understanding
of the hotspot observed in the TA data~\cite{Abbasi:2014lda}, which persists
after seven years of observation.

The upgrade of the Pierre Auger Observatory~\cite{Aab:2016vlz} includes installation
of a scintillator on top of each tank, improved electronics and underground muon
detectors.  With improved ability to separate the muon signal from the electromagnetic
signal in the surface detector, the upgraded array surface array will have sensitivity
to primary composition with 100\% duty cycle.  The goal is to determine the origin of
the flux suppression at the highest energy and to determine the feasibility of
astronomy with cosmic-ray protons. 

\begin{figure}[!t]
\centering
\includegraphics[width=7cm,clip]{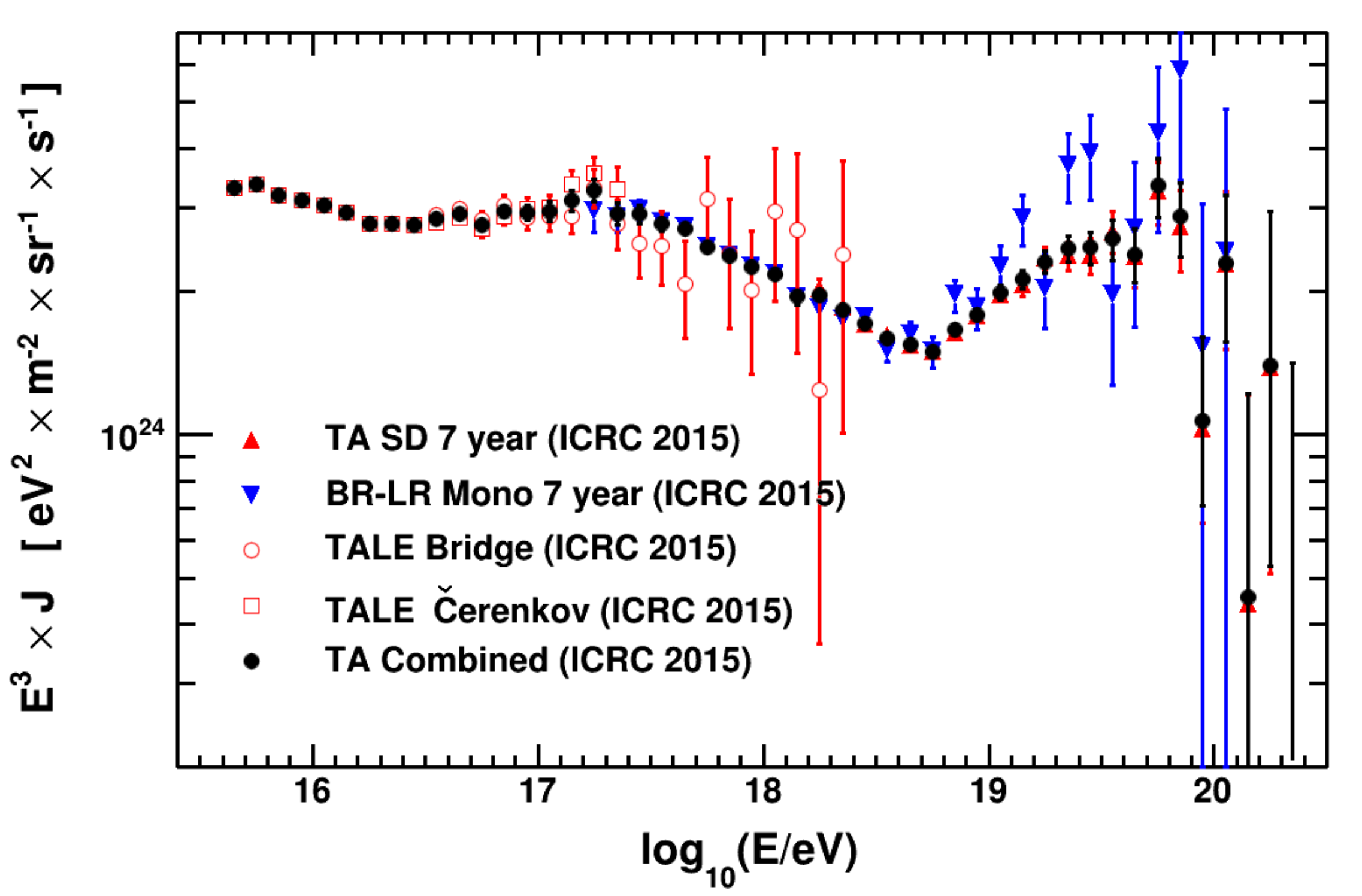}
\caption{The combined spectrum of TA presented at this conference~\cite{Matthews})
and shown here by kind permission of John Matthews.}
\label{fig:TA}       
\end{figure}

\subsection{Cosmic-ray overview}

Using its low energy extension (TALE), the TA group has
presented a combined spectrum that extends all the way down to 5~PeV (Fig.~\ref{fig:TA}).
The data are consistent with those shown in Fig.~\ref{fig:GER17} in the overlap
region.  The hardening of
the spectrum at $2\times 10^{16}$~eV and the second knee just above $10^{17}$~eV
are clearly visible, as well as the ankle and the suppression above $10^{20}$~eV.

\begin{figure}[!t]
\centering
\includegraphics[width=7cm,clip]{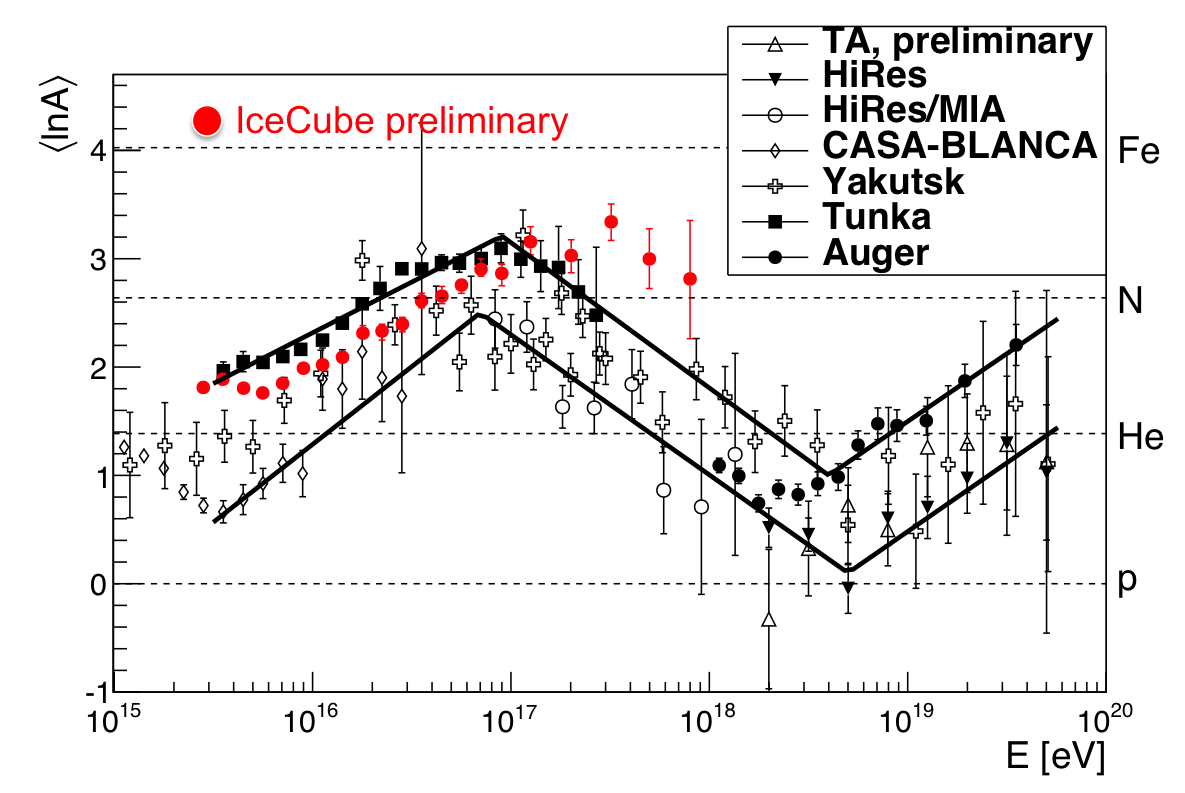}
\caption{Mean value of the natural logarithm of the primary mass
as a function of primary air shower energy.  The plot is from the paper
of Kampert \& Unger~\cite{Kampert:2012mx}), and the solid lines indicate
the ranged of data summarized there.  Preliminary composition data from
IceCube~\cite{Rawlins:2016bkc} are interpolated in the same way and added to
the Kampert \& Unger plot in red.
}
\label{fig:KampertUnger}       
\end{figure}

The composition also shows an energy dependence with
features that may be correlated with the spectrum,
as illustrated in Fig.~\ref{fig:KampertUnger}~\cite{Kampert:2012mx}. The plot is made by interpolating
the value of each measured depth of maximum between theoretical curves for protons and iron 
on a plot of $X_{max}$ vs energy (in this case using SIBYLL 2.1~\cite{Ahn:2009wx}).  The increasing
fraction of heavy nuclei up to $10^{17}$~eV, for example, can be interpreted naturally
as a rigidity-dependent effect as Galactic accelerators reach their maximum energy.
This is followed by an increase in the proton fraction that reaches a maximum
(minimum in $\langle\ell n(A)\rangle$) at the ankle.  The preliminary
IceCube result~\cite{Rawlins:2016bkc,Dembinski} is shown in red in Fig.~\ref{fig:KampertUnger}.
It agrees with the trend of the data up to $10^{17}$~eV, but remains at
a high level above that energy, although with large uncertainties.  

In his presentation Ptuskin~\cite{Ptuskin} reviewed the standard model of
cosmic-ray origin, with Galactic supernovae of different types producing the
cosmic-ray spectrum observed at Earth up to $\sim 10^{17}$~eV and an extra-galactic
component dominating above $10^{18}$~eV.  Analysis of low-energy iron isotopes~\cite{Hams:2015dkt}
shows that there must have been at least two nearby supernova remnants within
one or two million years of each other.  The general picture of supernovae in the disk
of the Galaxy exploding randomly and each accelerating cosmic rays for a period of time short
compared to the propagation time is explored  in Ref.~\cite{Blasi:2011fi}.  The
contribution of a particular supernova to the cosmic-ray flux
currently observed at Earth will be determined by the energy-dependent residence time
of cosmic-rays in the Galaxy coupled with 
time of the explosion and its distance from Earth.  Such considerations also 
have implications for the observed anistropy of cosmic rays~\cite{Blasi:2011fm},
which will reflect to some extent the particular history of supernova explosions in the Milky Way.

\begin{figure*}
\centering
\includegraphics[width=7cm,clip]{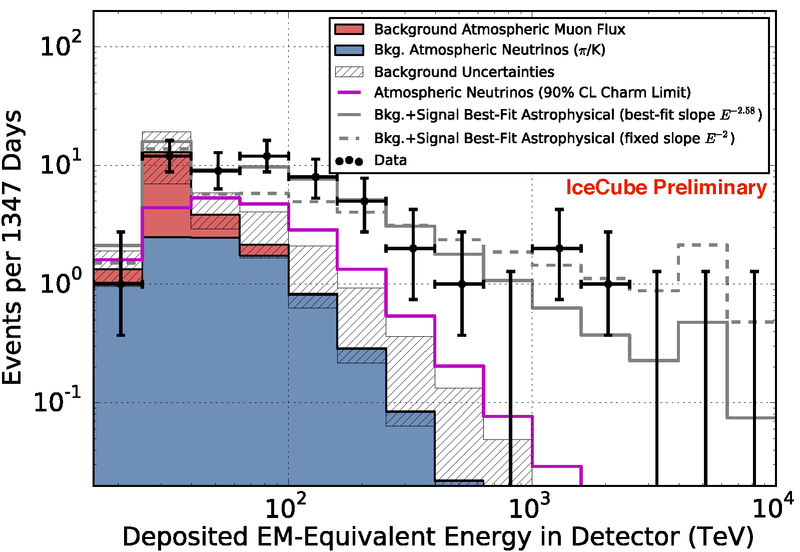},\,\,\includegraphics[width=8cm,clip]{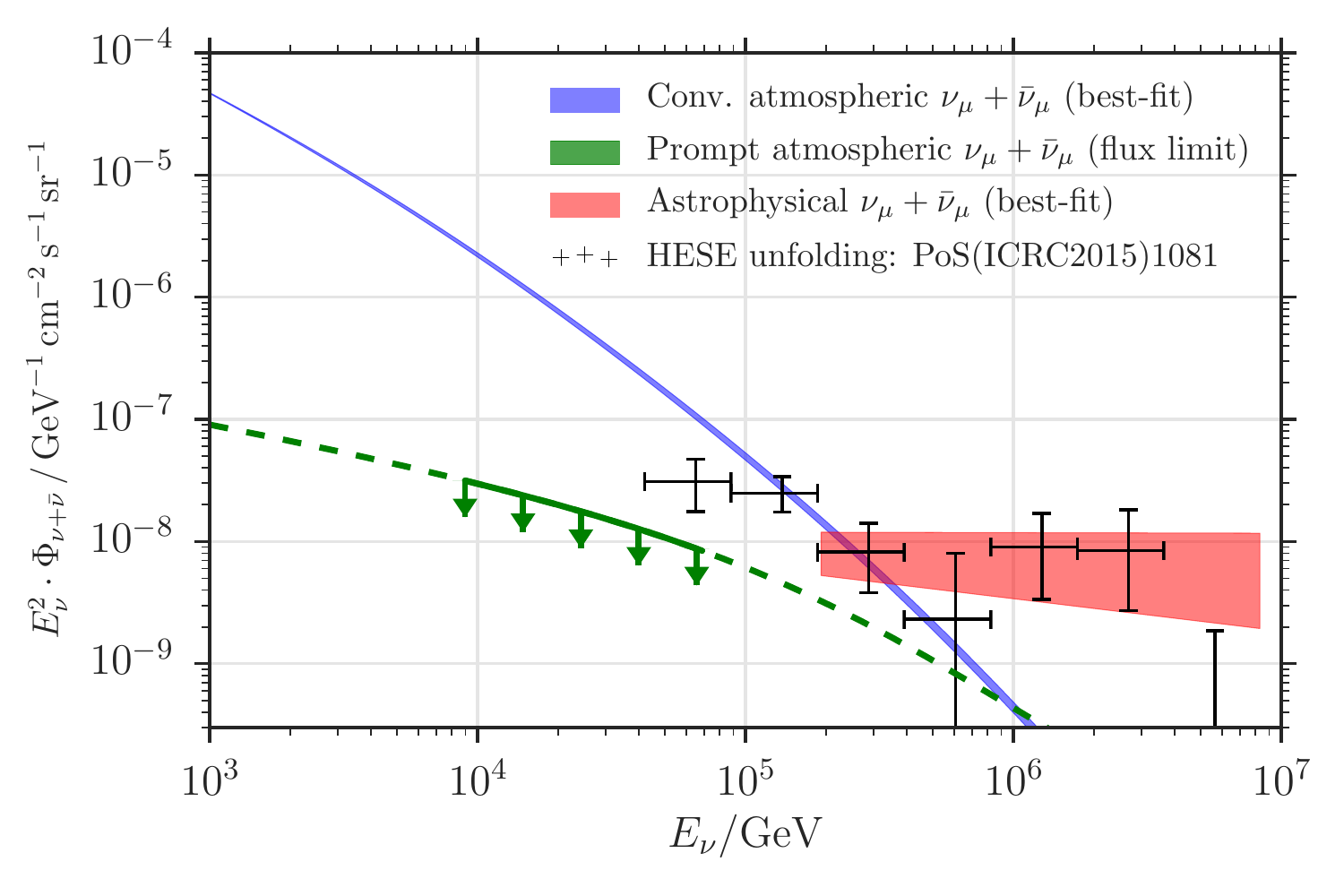}
\caption{Spectrum of high energy neutrinos from IceCube, Left: 4-year HESE analysis~\cite{Aartsen:2015zva};
Right: 6-year upward muon analysis~\cite{Aartsen:2016xlq}.}
\label{fig:HESE}
\end{figure*}

\section{Multi-messenger astronomy}
\label{sec-2}
Photons and neutrinos, being electrically neutral, propagate from their sources
in straight lines, unlike charged cosmic rays, which are bent in Galactic and cosmic
magnetic fields.  Therefore, to the extent that cosmic-rays interact in or near
their sources, gamma-rays and neutrinos should be good probes of cosmic-ray sources.
Neutrinos have the extra advantage of not being affected by interactions in transit (and the
corresponding disadvantage of being difficult to detect).  High-energy gamma-rays can be produced
electromagnetically, for example by inverse Compton scattering by electrons, as well
as from decay of $\pi^0$ produced in hadronic interactions.  Neutrinos can only be
produced by hadronic interactions.  Just as atmospheric neutrinos are produced by
interactions of cosmic rays in the atmosphere, any high-energy neutrino of extraterrestrial
origin will have been produced by the interaction of a proton (or less likely a nucleus)
somewhere in the cosmos.
The interaction may be with gas or by photo-pion production, but in both cases gamma-rays
would be produced through the corresponding neutral pion channel.  (The possibility
of $\gamma$-rays and neutrinos from dark matter is not considered here.)

\subsection{High-energy neutrinos in IceCube}

Figure~\ref{fig:HESE} shows the observation by IceCube of a hard spectrum of
astrophysical neutrinos extending above the steeply falling background
of atmospheric neutrinos.  The signal is seen consistently in two channels.
The high-energy starting event (HESE) analysis~\cite{Aartsen:2014gkd} requires the event to start
well inside the instrumented volume of IceCube in such a way that the outer
layers of optical modules exclude atmospheric muons from above, as well
as atmospheric neutrinos of sufficiently high energy to be 
accompanied by a muon~\cite{Schonert:2008is,Gaisser:2014bja}.  The astrophysical
signal can be identified to lower energy by increasing the veto
region and correspondingly reducing the size of the fiducial volume~\cite{Aartsen:2014muf}.
In the upward muon analysis~\cite{Aartsen:2016xlq},
the Earth is used as a shield to exclude atmospheric muons.  The signal then consists
mainly of neutrino-induced muons.  The astrophysical component is distinguished only
by its hard spectrum extending to high energy.

Accompanying each of the neutrino analyses is a list of the highest energy events with their locations
in the sky (right ascension and declination).  In the case of the HESE analysis,
the event type is given as cascade ($\nu_e$ and most $\nu_\tau$) or track ($\nu_\mu$).
The four-year sample includes 54 events with deposited energy of $30$~TeV and higher, 
of which fewer than twenty are background.
For the upward muon analysis, all events are tracks, and 
29 events with estimated muon energy $E_\nu > 200$~TeV are singled out as being
the most likely to be of astrophysical origin.  Sky maps are made for both cases, but
no source of high significance emerges.  The most energetic event in the upward muon
sample has a deposited energy in the detector of $2.6{\pm 0.3}$~PeV and an estimated muon energy
of $4.5$~PeV.  The probability
distribution for the parent neutrino energy peaks just below $10$~PeV~\cite{Aartsen:2016xlq}.

The spectral index that characterizes the astrophysical flux depends on the energy
range of the analysis, and the possibility of a break and or a cutoff in the spectrum
is not ruled out.  A combined maximum-likelihood analysis of the high-energy 
astrophysical neutrino flux measured with several IceCube analyses gives a value of
the energy flux of all neutrino flavors of
\begin{equation}
E_\nu^2\frac{{\rm d}N_\nu}{{\rm d}E_\nu}\approx 2\times 10^{-8} \,\,{\rm GeV\,cm}^{-2}{\rm sr}^{-1}\,{\rm s}^{-1}
\label{energy-flux}
\end{equation}
at $E_\nu=1$~PeV assuming the spectrum is a single power law \cite{Aartsen:2015knd}.  The fitted
differential spectral index, is $-2.5\pm~0.09$.  
It is interesting to note from
Fig.~\ref{fig:GER2_1} that the energy flux 
of cosmic rays has the same value at the ankle, which may be interpreted
as indicating that the energy flux of the extragalactic population of cosmic rays
is somehow related to the astrophysical neutrino flux, as suggested by
Waxman and Bahcall~\cite{Waxman:1998yy,Bahcall:1999yr}.  

In this context,
it is interesting to note that detection of cosmogenic neutrinos from photo-pion
production by interactions of ultra-high-energy cosmic rays with the CMB could
resolve the origin of the cutoff of the cosmic-ray spectrum at $100$~EeV.  A search
by IceCube for events of extremely high energy (EHE) with seven years of data~\cite{Aartsen:2016ngq}
finds no neutrinos with energy $> 10^7$~GeV.  This result places significant constraints
on some models of cosmogenic neutrinos.  Because neutrinos produced by interactions of
cosmic-rays with energy-per-nucleon $>10^9$~GeV would produce neutrinos with
$E_\nu>10^7$~GeV, the result also disfavors some models of astrophysical neutrinos 
associated with sources of ultra-high-energy cosmic rays. 

\subsection{Extragalactic source candidates}

The tables of high-energy neutrinos~\cite{Aartsen:2014gkd,Aartsen:2016xlq} constitute a small
fraction of the $\sim$~100,000 neutrinos per year reconstructed in IceCube.  Most of
these are atmospheric neutrinos.  Nevertheless, the most sensitive searches for 
point sources of astrophysical
neutrinos use the larger data samples that extend down to the TeV energy range.
Maximum likelihood techniques that take
account of the angular resolution and the deposited energy of each event are used to look for
point sources of neutrinos emerging from the smooth foreground of atmospheric neutrinos.
An all-sky analysis produces a sky map of significances.  In addition, a list of
sources selected, largely on the basis of gamma-ray observations, to be likely neutrino sources
is investigated.  The most sensitive search so far~\cite{Aartsen:2016oji} uses seven years of data and finds
no source of high significance in the all-sky analysis and places upper limits on
the targeted sources.  

The relation between a total observed intensity from all directions and the
number of events to be expected from nearby sources is addressed in general in~\cite{Lipari:2008zf}
and in the context of the IceCube discovery in~\cite{Ahlers:2014ioa}.
Upper limits on extragalactic sources, together with an
assumption about the cosmological evolution of the source class, lead to
a lower limit on the density of sources.  A recent analysis~\cite{Murase:2016gly} of the total
astrophysical intensity of neutrinos and the current upper limits of IceCube
suggests that the density of
extra-galactic sources should be $> 10^{-7}~(Mpc)^{-3}$.  Blazars and Gamma-Ray Bursts (GRB)
are among the rare, luminous sources that are being constrained by this line
of argument~\cite{Kowalski:2014zda}.  Moreover, there are experimental upper limits
from IceCube on the contribution to the IceCube signal of both blazars~\cite{Aartsen:2016lir}
and GRBs~\cite{Aartsen:2016qcr}.

Starburst galaxies are relatively abundant and have been proposed as a likely
source of neutrinos~\cite{Loeb:2006tw}.  The idea is that, because of the greater
rate of supernova explosions and the turbulence they generate,
there would be a higher intensity of cosmic rays produced and they would be confined
in the host galaxy sufficiently long that most protons would interact (calorimetric limit).
Since the target for the neutrino production is gas in the galaxy, the spectrum of
neutrinos and gamma-rays would extend down to low energy.  Fermi observations of
the diffuse gamma background~\cite{Ackermann:2014usa}
have been used to constrain the starburst model~\cite{Murase:2013rfa,Bechtol:2015uqb}.
In order to avoid producing more diffuse gamma-radiation than observed,
the spectra of cosmic-ray induced neutrinos and photons must have a
hard spectrum (differential spectral index $-\alpha$ with $\alpha < 2.2$~\cite{Murase:2013rfa}.
On the other hand, since most of the isotropic gamma-radiation comes from 
blazars~\cite{TheFermi-LAT:2015ykq}, it would be surprising if there were not
also neutrinos at some level~\cite{Resconi:2016ggj}.

Another possible neutrino source, discussed at this conference by Ptuskin~\cite{Ptuskin},
is Type IIn supernovae in external galaxies~\cite{Zirakashvili:2015mua}.  These are rare
core-collapse supernovae that accelerate protons into the dense wind of the progenitor
star.  The environment of the progenitor wind is sufficiently dense that the
SN expansion slows down on a time scale of 30 years.  Like the starburst
scenario, the target for neutrino (and gamma-ray) production is gas, so
the spectra extend to low energy and the Fermi isotropic gamma limit applies
(so the accelerated spectrum of protons must be hard).  A maximum energy of
$10^{17}$~eV/proton is found, which is sufficient to produce neutrinos up
to $\sim PeV$ (but not beyond).  Assuming a rate of 1\% of all core collapse
supernovae, a diffuse spectrum in rough agreement with the IceCube spectrum
is found.  Moreover, two tracklike events are in the direction of known
Type IIn supernovae.  HESE event 47 with 74 TeV deposited energy~\cite{Aartsen:2015zva} is
$1.35^\circ$ from SNIIn 2005bx and event 11 of the upward muon sample
with an estimated energy of $240$~TeV~\cite{Aartsen:2016xlq}
is $0.3^\circ$ from SNIIn 2005jq.

\subsection{Neutrinos and $\gamma$-rays from the Milky Way}

Several of the events in the high-energy samples come from near
the Galactic plane and could be of local origin in the Milky Way~\cite{Neronov:2015osa}.
In fact a local component is expected because
both neutrinos and photons are produced by cosmic-ray interactions of cosmic rays with interstellar gas 
during their diffusive propagation in the Galaxy.  In his paper on neutrinos,
Stecker~\cite{Stecker} pointed out that these neutrinos would have the same spectrum as that
of the primary cosmic rays ($\alpha\approx 2.7$).  At high energy the spectrum of
atmospheric neutrinos is one power steeper, so the Galactic neutrinos should become relatively
more prominent as energy increases.  In addition, the cosmic-ray spectrum in the central
region of the Galaxy could have a harder spectrum than observed locally at Earth~\cite{Gaggero:2015xza}.

The Galactic diffuse $\gamma$ emission is measured up
to $\approx 100$~GeV by Fermi~\cite{Ackermann:2012pya}.  It is concentrated along the 
Galactic ridge and is mostly from $\pi^0$-decay.  The ARGO-YBJ measurements of the diffuse
$\gamma$-emission up to a TeV are consistent with an extension
of the Fermi measurement~\cite{Bartoli:2015era}.  The corresponding spectrum
of $\nu_\mu+\bar{\nu}_\mu$ should have the same shape and a magnitude of
$2/3$ that of the gamma-rays at production and $1/3$ after oscillations~\cite{book}.
In the presentation at this
conference~\cite{diSciascio} the high intensity of gamma-rays from the
region of the Cygnus Cocoon~\cite{Ackermann:2011lfa} was also described.
This too should have a counterpart in neutrinos.  The ANTARES neutrino detector
has placed a limit~\cite{Adrian-Martinez:2016fei} on the intensity of multi-TeV neutrinos from the Galactic
plane that is approaching the higher prediction of Ref.~\cite{Gaggero:2015xza}.

Using gamma-ray data from Fermi-LAT as a template, predictions for Galactic neutrinos in 
the TeV to PeV range are examined in detail in Ref.~\cite{Ahlers:2015moa}.  The conclusion
is that less than 10\% of the high-energy events are of Galactic origin.  Limits are
also placed on the Galactic contributions in the TeV range.

\section{Summary and outlook}
\label{sec-3}

Understanding the origin of ultra-high energy cosmic rays and whether
their sources also produce neutrinos remains a key question~\cite{Aartsen:2016ngq}.
It is possible that the sources of the astrophysical neutrinos seen by
IceCube are not the sources of UHECR.  The current status of neutrinos and
cosmic-rays observed by IceCube is the subject of a recent review~\cite{Aartsen:2017von}.

To improve the ability to identify the sources of high-energy astrophysical
neutrinos, IceCube has started a real-time alert system~\cite{Aartsen:2016lmt}.  Its goal is
to maximize the ability to find electromagnetic counterparts of astrophysical
neutrinos seen in IceCube.  There are several types of alerts.  Single track-like events
(with good pointing and high energy) in the HESE and EHE samples generate alerts.
Alerts for follow-up by optical, X-ray and $\gamma$-ray detectors trigger on multiple events
from a single point in the sky in a limited time interval.

The discovery of high-energy astrophysical neutrinos motivates efforts to
accumulate more data with new and larger detectors.  Construction of KM3NeT
is starting, with the deployment of test lines at the Italian site, Capo Passero~\cite{Brunner}
and a Letter of Intent~\cite{Adrian-Martinez:2016fdl}.  
The proposal is to build a high-energy unit called ARCA
at Capo Passero and a densely instrumented array called ORCA at the French site near 
Marseille~\cite{Brunner}.  The first cluster of an expansion of Baikal to
a Gigaton Volume Detector (Baikal-GVD)~\cite{Avrorin:2013uyc} was installed in 2015.
Plans for a next generation IceCube-Gen2 are underway~\cite{Aartsen:2014njl,Aartsen:2015dkp}.
IceCube-Gen2 will include a densely instrumented subarray 
for neutrino physics~\cite{TheIceCube-Gen2:2016cap}.

The high-energy expansion of IceCube-Gen2 aims to increase the target volume by
an order of magnitude.  To reach the much larger volumes needed to detect
a large number of cosmogenic neutrinos, the radio technique is being pursued.
The first detectors of ARIANNA~\cite{Barwick:2016mxm} and of ARA~\cite{Allison:2015eky} 
prototypes are already in operation in Antarctica.  A review of the current status
of high-energy neutrino astronomy that covers current and future detectors and
the science they will address is in preparation~\cite{book2}.

\section*{Acknowledgment}  I thank John Matthews for permission to use his composite
spectrum of Telescope Array and TALE (Fig.~\ref{fig:TA}), and I thank Javier Gonzalez 
for adding the IceCube data points in Fig.~\ref{fig:KampertUnger}.  I thank Christopher Wiebusch
for helpful comments on a draft version of this paper, and I acknowledge support from
the U.S. National Science Foundation (PHY-1505990).

%
%

\end{document}